\documentclass[aps,amsmath,amssymb,twocolumn,twoside,floatfix,preprintnumbers,nobibnotes,nofootinbib]{revtex4}

\usepackage{docs}
\usepackage{epsfig}
\expandafter\ifx\csname package@font\endcsname\relax\else
 \expandafter\expandafter
 \expandafter\usepackage
 \expandafter\expandafter
 \expandafter{\csname package@font\endcsname}%
\fi
\DeclareRobustCommand\substyle{\name@idx{document substyle}}%
\DeclareRobustCommand\classoption{\name@idx{document class option}}%
\DeclareRobustCommand\classname{\name@idx{document class}}%
\def\name@idx#1#2{%
 {\ttfamily#2}%
 \index{#2\space#1=\string\ttt{#2}\space#1}\index{#1>#2=\string\ttt{#2}}%
}%

\def\kpn{K^+\rightarrow\pi^+\nu\bar\nu}

\def\klpn{K_{ L}\rightarrow\pi^0\nu\bar\nu}

\newcommand{\lsim}{
\mathrel{\hbox{\rlap{\hbox{\lower4pt\hbox{$\sim$}}}\hbox{$<$}}}}

\newcommand{\gsim}{
\mathrel{\hbox{\rlap{\hbox{\lower4pt\hbox{$\sim$}}}\hbox{$>$}}}}

\def\eps{\varepsilon}
\def\epe{\varepsilon'/\varepsilon}

\newcommand{\Heff}{{\cal H}_\text{ eff}}

\DeclareMathOperator{\im}{Im}

\newcommand{\be}{\begin{equation}}
\newcommand{\ee}{\end{equation}}
\newcommand{\bea}{\begin{eqnarray}}
\newcommand{\eea}{\end{eqnarray}}
\newcommand{\nn}{\nonumber}
\newcommand{\bi}{\begin{itemize}}
\newcommand{\ei}{\end{itemize}}
\newcommand{\ord}{{\cal O}}

\begin{document}

\preprint{TUM-HEP-714/09, MPP-2009-25}

\title{\boldmath Insights from the Interplay of $K\to\pi\nu\bar\nu$ and $\eps_K$ 
on the New Physics Flavour Structure}%
\author{Monika Blanke}%
\affiliation{Physik Department, Technische Universit\"at M\"unchen, D-85748 Garching, Germany\\
Max-Planck-Institut f{\"u}r Physik (Werner-Heisenberg-Institut), 
 D-80805 M{\"u}nchen, Germany}

\begin{abstract}
In certain new physics (NP) models, such as the Littlest Higgs model with T-parity, a strict correlation between the  $\klpn$ and $\kpn$ branching ratios has been observed, allowing essentially only for two branches of possible points, while in other NP frameworks, such as the general MSSM or warped extra dimensional models, no visible correlation appears. We analyse the origin of the correlation in question
and show it to be a direct consequence of the stringent experimental constraint on $\eps_K$, provided that the NP enters with comparable strength and a universal weak phase in both $\Delta S=2$ and $\Delta S=1$ transitions. This happens in many NP scenarios with either only SM operators, or where the NP induces exclusively right-handed currents while the left-right $\Delta S=2$ operators are absent.  On the other hand, if the NP phases in $\Delta S=2$ and $\Delta S=1$ processes are uncorrelated, $\eps_K$ has no power to put constraints on the $K\to\pi\nu\bar\nu$ system. The latter appears in particular in those NP models where $K^0-\bar K^0$ mixing receives contributions from the chirally enhanced left-right operators. We discuss the stability of the correlation in question against small deviations from the assumption of universal $\Delta S=2$ and $\Delta S=1$ weak phases, and in the presence of non-negligible NP contributions to $\eps_K$.
\end{abstract}

\maketitle

\section{Introduction}

The $K\to\pi\nu\bar\nu$ decays, being theoretically very clean and extremely suppressed in the Standard Model (SM), are known to be one of the best probes of new physics (NP) in the flavour sector. Recent reviews of these decays both in and beyond the SM can be found in \cite{Buras:2004uu}, here we just quote for completeness the presently available SM predictions, obtained at the NNLO level \cite{Buras:2005gr}
\be\label{eq:KLSM}
Br(\klpn)_\text{SM}= (2.76\pm 0.40)\cdot 10^{-11} 
\ee
and
\be\label{eq:KpSM}
Br(\kpn)_\text{SM}=(8.5 \pm 0.7) \cdot 10^{-11}  \,.
\ee

Unfortunately the $K\to\pi\nu\bar\nu$ decays are experimentally very challenging, so that for $Br(\klpn)$ only an upper bound \cite{Ahn:2007cd}
\be
Br(\klpn)_\text{exp}<6.7\cdot 10^{-8}\qquad (90\% \text{\,C.L.})
\ee
is available, while the present measurement of $Br(\kpn)$ \cite{Artamonov:2008qb}
\be\label{eq:kpnex}
Br(\kpn)_\text{exp}=(17.3^{+11.5}_{-10.5})\cdot 10^{-11}
\ee
is still plagued by large uncertainties.

While observing one day these two branching ratios outside the ranges \eqref{eq:KLSM}, \eqref{eq:KpSM} predicted in the SM would clearly be a spectacular sign of new physics (NP), it is even more interesting to consider both decay rates simultaneously. In fact, while in some NP models like the Littlest Higgs model with T-parity (LHT) \cite{Blanke:2006eb,Goto:2008fj,LHTupdate} or the minimal 3-3-1 model \cite{Promberger:2007py} a stringent correlation in the $K\to\pi\nu\bar\nu$ system has been found, in other NP frameworks like the general MSSM  \cite{Nir:1997tf,Isidori:2006qy} or Randall-Sundrum (RS) models with bulk fields \cite{Blanke:2008yr,Haisch-BF08} any values of the decay rates in question consistent with the model-independent Grossman-Nir (GN) bound \cite{Grossman:1997sk} appear possible.

Stimulated by this observation, in the present paper we aim to reveal the origin of the  correlation in question and analyse the conditions under which it appears. To this end let us briefly recall certain properties of the models in question:
\bi
\item Both the LHT model and the minimal 3-3-1 model contain new sources of flavour and CP-violation and in particular new CP-violating phases \cite{Hubisz:2005bd,Blanke:2006sb,Promberger:2007py}. While in the minimal 3-3-1 model flavour transitions in a given meson system are governed by a single weak phase, in the LHT model a priori various contributions with different weak phases are present.  On the other hand it is common to both models that no new flavour violating operators beyond the ones already present in the SM appear.
\item
In the general MSSM and in RS models with bulk fields, both with and without custodial protection, flavour changing neutral current (FCNC) processes are mediated by new operators in addition to the usual SM left-handed ones \cite{Csaki:2008zd,Blanke:2008zb,Agashe:2008uz,Blanke:2008yr}. Also in these models new sources of flavour and CP-violation are present.
\ei

This observation raises the suspicion that the correlation in the $K\to\pi\nu\bar\nu$ system in question could be a remnant of the absence of new flavour violating operators. Indeed various probes of the NP operator structure through correlations between various rare decay rates have been discussed extensively in the literature. A unique probe of new flavour violating operators is given by the $B_{s,d}\to\mu^+\mu^-$ decays, that can only receive large enhancements by scalar operator contributions. Proposals to test the presence of new flavour violating operators have also been made using the $K_L\to\pi^0\ell^+ \ell^-$ decays \cite{Isidori:2004rb}, the correlation between the $K^+\to\pi^+\nu\bar\nu$ and $K_L\to\mu^+\mu^-$ decay rates \cite{Blanke:2008yr} or the three body leptonic $\mu$ and $\tau$ decays, by comparing their branching ratios to the ones of $\mu\to e\gamma$ and $\tau\to\ell\gamma$ \cite{Ellis:2002fe} or by performing a Dalitz plot analysis \cite{Dassinger:2007ru}.

However, the situation in the $K\to\pi\nu\bar\nu$ system is peculiar, as the relevant effective Hamiltonian contains only the operators
\be
(\bar sd)_{V-A} (\bar\nu\nu)_{V-A}\,,\qquad
(\bar sd)_{V+A} (\bar\nu\nu)_{V-A}\,.
\ee
Furthermore, as both $K$ and $\pi$ are pseudoscalar mesons, effectively only the linear combination 
\be\label{eq:operator}
(\bar sd)_V (\bar\nu\nu)_{V-A}
\ee
contributes. Therefore the correlation in question can clearly {\it not} be a result of various contributions adding up in different ways in $Br(\kpn)$ and $Br(\klpn)$, as happens in most other cases mentioned above. Consequently the correlation within the $K\to\pi\nu\bar\nu$ system can {\it not} be induced by the NP operator structure in $\Delta S=1$ transitions, which has to be tested by other means \cite{Isidori:2004rb,Blanke:2008yr}.

On the other hand, the $K\to\pi\nu\bar\nu$ decays offer an excellent probe of the weak phase appearing with the operator in \eqref{eq:operator} \cite{Buras:2004ub}. While $\kpn$, being a CP-conserving decay, is sensitive to the absolute value of the Wilson coefficient of \eqref{eq:operator}, the CP-violating $\klpn$ decay measures its imaginary part.
As generally the NP phase in $\Delta S=1$ is arbitrary and not yet constrained by the data,\footnote{While there exist rather precise data on direct CP-violation by means of the parameter $\epe$, in this case the SM prediction is unfortunately only poorly known so that no useful constraint on NP can be obtained. See \cite{Buras:2003zz} for a detailed discussion of the present situation in the SM.} the strict correlation in the $K\to\pi\nu\bar\nu$ system observed in some NP models must be due to other FCNC constraints.

Indeed in what follows we will demonstrate that in models in which
NP in $\Delta S=2$ and $\Delta S=1$ transitions is correlated to each other, i.\,e.\ the CP-violating phases are equal in both cases (apart from a trivial factor 2) and the NP amplitudes are of comparable relative size,
the stringent experimental constraint on CP-violation in $K^0-\bar K^0$ mixing, parameterised by $\eps_K$, constrains the $K\to\pi\nu\bar\nu$ decay rates to lie within the two branches observed within the LHT model \cite{Blanke:2006eb,LHTupdate}.

We will see that the precise shape of the correlation in question depends on the SM prediction for $\eps_K$ in comparison with the data. While at present the SM appears to reproduce the data within present uncertainties, in particular if the so far relatively imprecise tree level value for the CKM angle $\gamma$ is used, recent studies \cite{Buras:2008nn,Buras:2009pj,Lunghi:2008aa} hint for the possibility that the SM cannot account for the measured value of $\eps_K$ and a NP contribution of roughly $+20\%$ is required. As the situation is not conclusive at present, in our analysis we will consider two scenarios:
\begin{enumerate}
\item
The SM is in good agreement with the data on $\eps_K$, and the NP contributions are allowed to amount to at most $\pm5\%$ of the SM contribution.
\item
The SM alone cannot reproduce the measured value of $\eps_K$, but a $\sim +20\%$ NP contribution is required. Again in order to account for unavoidable theoretical parametric uncertainties, we take the NP contribution to be $(20\pm5)\%$ of the SM contribution.
\end{enumerate}
Fortunately, the theoretical knowledge of $\eps_K$ will improve significantly in the coming years, thanks to further improved lattice determinations of $\hat B_K$ and more precise measurements of the CKM parameters $|V_{cb}|$ and $\gamma$ at future facilities. Therefore when eventually the $K\to\pi\nu\bar\nu$ decays will be measured with sufficient precision to test the correlation in question, we will already know which of the above scenarios is satisfied in nature, and a $\pm5\%$ uncertainty in the SM prediction for $\eps_K$, while constituting an optimistic scenario, could be achieved at that stage. In any case the correlation in question does not depend qualitatively on this assumption.

If $\eps_K$ will indeed turn out to be well described by the SM prediction, finding the $K\to\pi\nu\bar\nu$ branching ratios outside the correlation in question would lead us to the insight that $\Delta S=2$ and $\Delta S=1$ transitions are {\it not} strongly correlated. In particular the following possibilities appear:
\begin{enumerate}
\item
$\Delta S=2$ and $\Delta S=1$ transitions are {\it not} governed by a universal weak phase. The origin of such a non-universality could be
\bi
\item the presence of various contributions with different weak phases, affecting $\Delta S=2$ and $\Delta S=1$ transitions in a different manner,
\item 
the presence of new operators (in particular the chirally enhanced left-right ones) in $\eps_K$ that spoil the direct correspondence between $\Delta S=2$ and $\Delta S=1$ physics. 
We note however that if NP effects are dominantly induced by right-handed currents that generate only $(V+A)\otimes(V+A)$ contributions to $\eps_K$, the correlation in question is still present.
\ei
\item
NP effects in $\Delta S=1$ transitions are significantly enhanced over the corresponding $\Delta S=2$ effects. This can appear e.\,g.\ in certain regions of the MSSM parameter space, where the squarks are much lighter than the gauginos \cite{Isidori:2006qy}. 
\end{enumerate}

On the other hand, if $\eps_K$ will indeed turn out to be significantly affected by NP contributions, the correlation in question is weakend and its use is shifted towards a better understanding of models with a universal phase in $\Delta S=2$ and $\Delta S=1$ transitions. In this case the measurement of the $K\to\pi\nu\bar\nu$ decay rates can be used to precisely determine the relative size of NP amplitudes in $K^0-\bar K^0$ mixing and the $K\to\pi\nu\bar\nu$ decays within this specific NP scenario.

The paper is organised as follows. In Section \ref{sec:frame} we introduce the general framework to describe NP contributions to the $\eps_K$ parameter and to the $K\to\pi\nu\bar\nu$ decays. 
Section \ref{sec:uniphase} is devoted to the discussion of models with a universal weak phase in $\Delta S=2$ and $\Delta S=1$ processes. In Section \ref{sec:noNP} we consider as a simple analytic example the case in which $\eps_K$ is NP-free and show that the allowed range in the $K\to\pi\nu\bar\nu$ plane reduces then to two straight lines. In Section \ref{sec:smallNP} we study the more realistic case of a possible small NP contribution to $\eps_K$, still finding a clear 2-branch correlation in the $K\to\pi\nu\bar\nu$ system, provided that the $\Delta S=1$ NP amplitude is not significantly enhanced over the $\Delta S=2$ one. In Section \ref{sec:largeNP} we consider the case of relevant NP contributions to $\eps_K$, implying a partial loss of correlation depending on the relative sizes of $\Delta S=2$ and $\Delta S=1$ amplitudes.
Subsequently in Section  \ref{sec:nonuniphase} we  soften the assumption of a universal phase by allowing for a second contribution with arbitrary phase to $\Delta S=1$ transitions. We will see that the correlation of the $K\to\pi\nu\bar\nu$ decays gets partly lost in this case. Still, we find that under the  assumption that this new contribution is suppressed with respect to the one entering also $\Delta S=2$ transitions, the correlation found previously is still present, albeit weaker.
In Section \ref{sec:newop} we briefly discuss the impact of the NP operator structure on the correlation in question. In models with only SM operators $\Delta S=2$ and $\Delta S=1$ are often strongly correlated, which we show explicitly for the Littlest Higgs model with T-parity. We then show that the situation is analogous for models inducing only right-handed currents. The situation is drastically different in the presence of the chirally enhanced left-right operator contributions to $K^0-\bar K^0$ mixing, and the correlation in the $K\to\pi\nu\bar\nu$ system is completely lost in that case.
 These observations can thus in principle be used to test the operator structure of NP once both $Br(\kpn)$ and $Br(\klpn)$ will be measured with sufficient precision.
We summarise our findings in Section \ref{sec:sum}.

\section{Basic Framework}\label{sec:frame}

\subsection{Preliminaries}

For the model-independent discussion of NP contributions to $K^0-\bar K^0$ mixing and the $K\to\pi\nu\bar\nu$ decays, it will turn out to be useful to work with the parameterisation introduced in what follows.

\subsection{\boldmath $\eps_K$ in the Presence of New Physics}

The off-diagonal mixing amplitude $M_{12}^K$ governing $K^0-\bar K^0$ oscillations can generally be written as
\begin{equation}
M_{12}^K=\frac{G_F^2}{12\pi^2}F_K^2\hat
B_K m_K M_{W}^2\overline{M_{12}^K}\,,
\end{equation}
where $\overline{M_{12}^K}$ can generally be divided into a SM and a NP part,
\be
\overline{
  M_{12}^K}=  \left(\overline{M_{12}^K}\right)_\text{SM}+  \left(\overline{M_{12}^K}\right)_\text{NP}\,.
\ee

The SM contribution is given by
\bea
\left(\overline{
  M_{12}^K}\right)_\text{SM}&=&\left(\lambda_c^{(K)*}\right)^2\eta_1S_c+\left(\lambda_t^{(K)*}\right)^2\eta_2S_t\nn\\
&&{}+2\lambda_c^{(K)*}\lambda_t^{(K)*}\eta_3S_{ct}\,.
\eea
Here, $\lambda_c^{(K)}=V_{cs}^*V_{cd}$ and $\lambda_t^{(K)}=V_{ts}^*V_{td}$ are the relevant CKM factors, $\eta_1,\eta_2,\eta_3$ are QCD corrections evaluated at the NLO level in \cite{Buras:1990fn}, and $S_c, S_t, S_{ct}$ are the SM one-loop functions.

Any kind of NP contribution to $\overline{M_{12}^K}$ can be parameterised in terms of its amplitude $R_{\Delta S=2}$ and its CP-violating phase $\phi_{\Delta S=2}$ as
\be\label{eq:M12NP}
\left(\overline{
  M_{12}^K}\right)_\text{NP}= \left(R_{\Delta S=2} \,e^{-i\phi_{\Delta S=2}}\right)^2\,,
\ee
where the square will turn out to be useful later on.

The parameter $\eps_K$, measuring the amount of mixing induced CP-violation in the $K\to\pi\pi$ decays, can then be written as
\be
\eps_K = \kappa_\eps e^{i\phi_\eps} \frac{\im M_{12}^K}{\Delta M_K}\,,
\ee
where experimentally $\phi_\eps=(43.51\pm0.05)^\circ$, and the parameter $\kappa_\eps =0.92\pm0.02$ in the SM \cite{Buras:2008nn} and under mild assumptions also in the presence of NP \cite{Buras:2009pj}. 

The experimental value $|\eps_K| =(2.229\pm 0.012)\cdot 10^{-3}$ turns out to be  somewhat larger \cite{Buras:2008nn,Buras:2009pj} than the SM prediction\footnote{See \cite{Lunghi:2008aa} for an alternative discussion.}, albeit still compatible due to the uncertainty mainly in $|V_{cb}|$ and the non-perturbative parameter $\hat B_K$, and in the tree level determination of the CKM angle~$\gamma$.

\subsection{\boldmath $K\to\pi\nu\bar\nu$ in the Presence of New Physics}

The $K\to\pi\nu\bar\nu$ decay rates can very generally be written as \cite{Buras:2004ub,Blanke:2006eb,Blanke:2008yr}
\bea\label{eq:BrKp}
Br(\kpn) &=& \kappa_+ \Big(\tilde r^2 A^4 R_t^2 |X_K|^2 \nn\\
&&\hspace*{-20mm} + 
     2 \tilde r \bar P_c(x) A^2 R_t |X_K| \cos\beta^K_X + \bar P_c(x)^2\Big)\,,\\
Br(\klpn) &=& \kappa_L \tilde r^2 A^4 R_t^2  |X_K|^2\sin^2\beta_X^K\,,\label{eq:BrKL}
\eea
with
\bea
X_K &=& X_0(x_t) + \frac{1}{\lambda_t^{(K)}}R_{K\to\pi\nu\bar\nu} e^{i\phi_{K\to\pi\nu\bar\nu}}\nn\\
&\equiv& |X_K| e^{i\theta_X^K}\,,\\
\beta_X^K &=& \beta-\beta_s-\theta_X^K\,.
\eea
Here $X_0(x_t)$ is the SM loop function describing the $s\to d\nu\bar\nu$ transition, and the NP contribution is parameterised by its amplitude $R_{K\to\pi\nu\bar\nu}$ and its weak phase $\phi_{K\to\pi\nu\bar\nu}$.
We note that due to the special structure of the $K\to\pi\nu \bar\nu$ decays, essentially only the operator
\be
(\bar sd)_V (\bar\nu\nu)_{V-A}
\ee
enters both branching ratios, so that the NP contributions to $\kpn$ and $\klpn$ are strongly correlated and can be parameterised by the single function $X_K$. Furthermore \cite{Mescia:2007kn,Buras:2005gr},
\begin{gather}
\kappa_+=(5.36\pm0.03)\cdot 10^{-11}\,, \quad \kappa_L=(2.31\pm0.01)\cdot 10^{-10}\,,\\
\tilde r = \left|\frac{V_{ts}}{V_{cb}}\right|\,,\qquad \bar P_c(x)= \left(1-\frac{\lambda^2}{2}\right)(0.42\pm 0.05) \,,\\
A= \frac{\left|V_{cd}V_{cb}^*\right|}{\lambda^3}\,,\qquad R_t= \left| \frac{V_{td}V_{tb}^*}{V_{cd}V_{cb}^*}\right|\,,\\
\beta=-\text{arg}(V_{td})\,, \qquad \beta_s= -\text{arg}(-V_{ts})\,.
\end{gather}

In our numerical analysis we have used for the values of the relevant input parameters the ones collected in Table 2 of \cite{Blanke:2008yr}.


\boldmath
\section{A Universal Weak Phase in $K$ Physics}
\unboldmath\label{sec:uniphase}

\subsection{Preliminaries}

Let us first consider the simple scenario in which NP CP-violation enters $\Delta S=2$ and $\Delta S=1$ processes in a universal manner, i.\,e.
\be\label{eq:uni}
\phi_{\Delta S=2} = \phi_{K\to\pi\nu\bar\nu}\equiv \phi\,.
\ee
While this assumption may seem an ad hoc one due to the a priori different structures of particle-antiparticle mixing and rare decays, it generally arises if only one NP source of flavour violation enters both $\Delta S=2$ and $\Delta S=1$ physics. As for the process of $K^0-\bar K^0$ mixing a double $(s\to d)\otimes (s\to d)$ transition is required, while the $K\to\pi\nu\bar\nu$ decays are mediated by a single $(s\to d)$ transition, in this case the weak phases will differ only by a factor of two, corresponding to the square in \eqref{eq:M12NP}.

Indeed there exist NP models of this type, e.\,g.\ the minimal 3-3-1 model \cite{Ng:1992st,Promberger:2007py} or the next-to-minimal flavour violating class of models \cite{Agashe:2005hk}, provided no new operators are present.

\subsection{\boldmath Assuming no NP in $\eps_K$ -- an Analytic Exercise}\label{sec:noNP}

In order to get a notion for the implications of the universality assumption \eqref{eq:uni}, let us start by considering the case that no NP appears in $\eps_K$. In order to achieve this
\be\label{eq:Im=0}
\im \left(M_{12}^K\right)_\text{NP} = 0
\ee
is required, implying
\be\label{eq:phi=npi2}
\phi = n \frac{\pi}{2}\qquad (n=0,1,2,3)\,.
\ee
Due to the experimental, parametric and theoretical uncertainties entering $\eps_K$
we will of course never know whether \eqref{eq:Im=0} is exactly satisfied in nature. Still it is useful to first consider this simplified toy scenario. As this case can easily be treated analytically, we will get a better understanding of how the 2-branch correlation in the $K\to\pi\nu\bar\nu$ system emerges. We note that eqs.\ \eqref{eq:Im=0} and \eqref{eq:phi=npi2} are phase convention dependent and valid only in the standard phase conventions for the CKM matrix. While 
\eqref{eq:Im=0} and \eqref{eq:phi=npi2} would look different if different phase conventions were chosen, the resulting constraints on the $K\to\pi\nu\bar\nu$ are of course independent of this choice.

Inserting the possible solutions for $\phi$ in \eqref{eq:phi=npi2} into eq.\ \eqref{eq:BrKp}, \eqref{eq:BrKL} and writing $Br(\klpn)$ as a function of $Br(\kpn)$, we find
\be\label{eq:b1}
Br(\klpn) = Br(\klpn)_\text{SM}
\ee
for $n=0,2$ independent of the value of $Br(\kpn)$, i.\,e.\ a horizontal line in the $Br(\kpn)-Br(\klpn)$ plane. As in this case the NP contribution to the $K\to\pi\nu\bar\nu$ decay amplitude is real, it has no impact on $\klpn$ being a CP-violating mode but affects only the CP-conserving $\kpn$ decay.

\begin{figure}
\epsfig{file=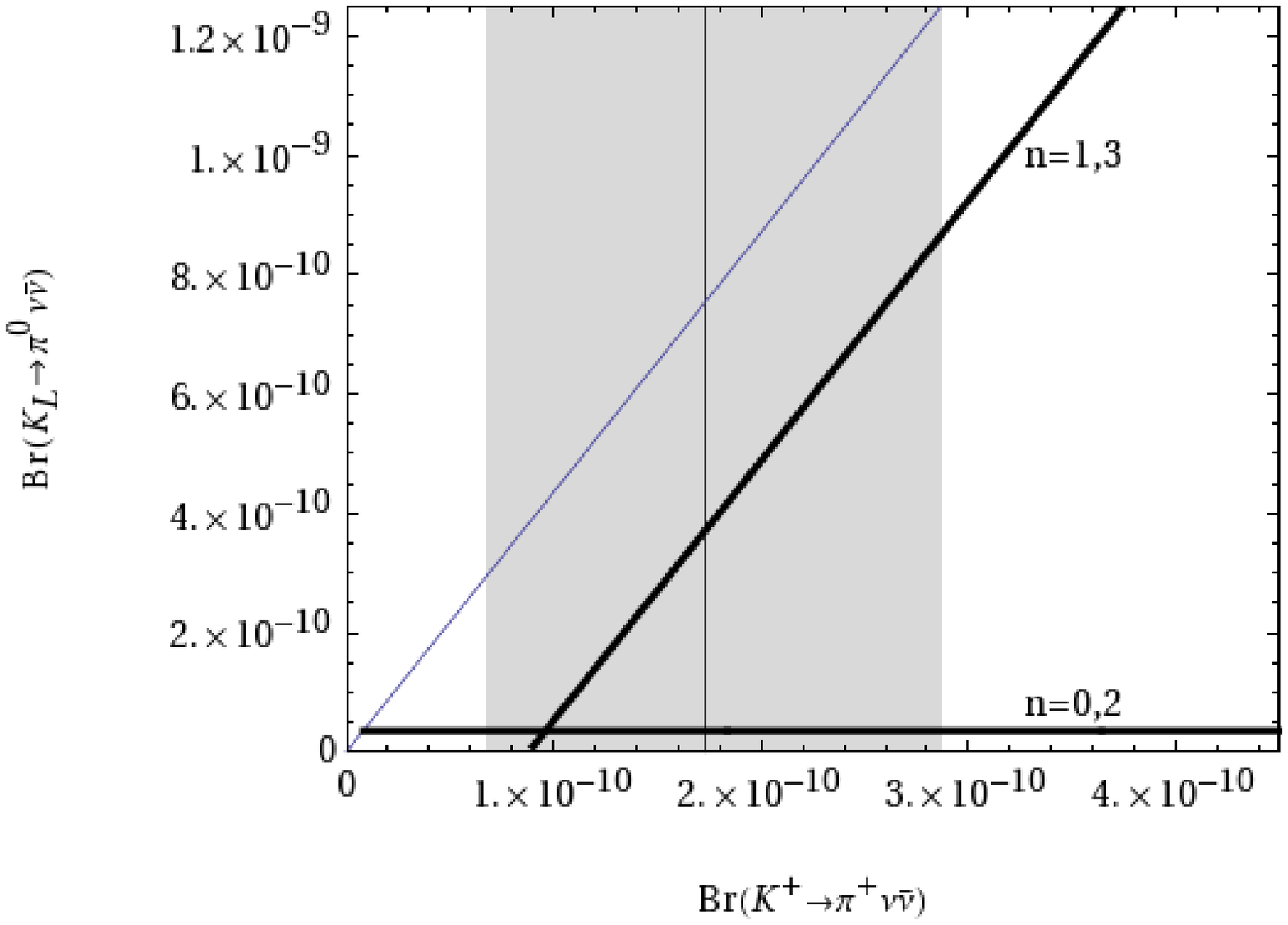,scale=.4}
\caption{$Br(\klpn)$ as a function of $Br(\kpn)$ in the scenario $\phi = n {\pi}/{2}$ $(n=0,1,2,3)$. The thin blue line shows the GN-bound, while the experimental $1\sigma$ range for $Br(\kpn)$ is displayed by the grey band.\label{fig:branches}}
\end{figure}

 For $n=1,3$ instead the NP contribution is purely imaginary, so that $Br(\klpn)$ is maximally affected. Solving then eqs.\ \eqref{eq:BrKp}, \eqref{eq:BrKL} for $Br(\klpn)$, we find
\bea\label{eq:b2}
Br(\klpn) &=& \frac{\kappa_L}{\kappa_+}Br(\kpn) \nn\\
&&-\kappa_L\left[\tilde r^2 A^4 R_t^2 X_0(x_t)^2\cos^2(\beta-\beta_s)\right.\nn\\
&&  + 2 \tilde r \bar P_c(x) A^2 R_t X_0(x_t) \cos(\beta-\beta_s)\nn\\
&&\left. +\bar P_c(x)^2 \right]\,.
\eea
This relation is represented by a straight line in the $Br(\kpn)-Br(\klpn)$ plane  parallel to the Grossman-Nir (GN) bound \cite{Grossman:1997sk}
\be
Br(\klpn) \le \frac{\kappa_L}{\kappa_+}Br(\kpn)\,,
\ee
but shifted downwards due to the subtrahend in \eqref{eq:b2}, so that it crosses the SM prediction. We note that the slope of this second branch, similarly to the one of the GN-bound, does not depend on any model-specific assumptions, but is a universal prediction for purely imaginary NP contributions to the $K\to\pi\nu\bar\nu$ system.

The two branches \eqref{eq:b1} and \eqref{eq:b2} are shown in Fig.\ \ref{fig:branches}. Their crossing point indicates the SM predictions for $Br(\kpn)$ and $Br(\klpn)$. We note that the only uncertainties in the precise position of the two branches under consideration arise from the SM predictions for the $K\to\pi\nu\bar\nu$ decay rates and the small parametric uncertainties in $\kappa_L$ and $\kappa_+$.

\subsection{\boldmath Small NP Effects in $\eps_K$}\label{sec:smallNP}

Unfortunately, due to the experimental, parametric and theoretical uncertainties in the determination of $\eps_K$, we will never know whether $\eps_K = (\eps_K)_\text{SM}$ is exactly satisfied in nature. While at present the uncertainties in the SM prediction for $\eps_K$ are still sizable, mainly due to the errors in the tree level determinations of $|V_{cb}|$ and $\gamma$,
further improved lattice determinations of $\hat B_K$ and more precise measurements of the relevant CKM parameters at future facilities will improve the situation significantly before the $K\to\pi\nu\bar\nu$ branching ratios will precisely be measured.

\begin{figure}
\epsfig{file=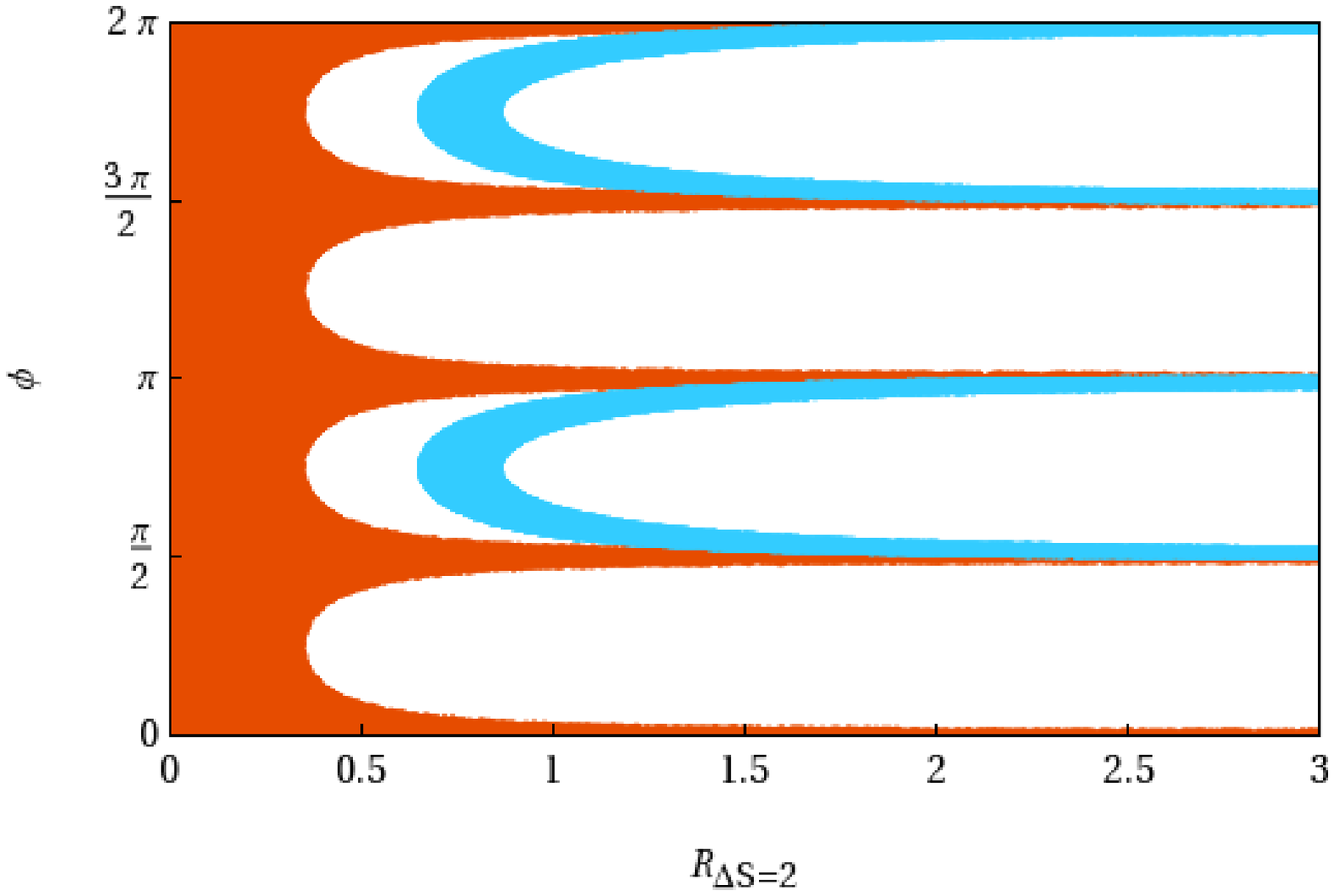,scale=.4}
\caption{Allowed ranges in the $(R_{\Delta S=2},\phi_{\Delta S=2})$ plane. Orange (darker grey) area: assuming a NP contribution to $\eps_K$ of at most $\pm 5\%$ of the SM contribution; blue (lighter grey) area: assuming a NP contribution to $\eps_K$ of  $+(20\pm 5)\%$ of the SM contribution.\label{fig:eps-1}}
\end{figure}

Therefore rather than including the effect of the {\it present} uncertainties in $\eps_K$ into our analysis, we will now assume a {\it future} accuracy of $5\%$ for the SM prediction and a good agreement with the data.\footnote{The case in which the SM cannot account for the measured value of  $\eps_K$, as hinted at in \cite{Buras:2008nn,Buras:2009pj,Lunghi:2008aa}, will be discussed in Section \ref{sec:largeNP}.} More precisely, we allow for a NP contribution to $\eps_K$ of at most $\pm 5\%$ of the SM contribution.\footnote{While a future $5\%$ accuracy of $(\eps_K)_\text{SM}$ may seem optimistic, we would like to stress that the correlation pointed out here does not depend crucially on this assumption -- rather it should be considered as a numerical example.} The impact of this constraint on the allowed parameter space in the $(R_{\Delta S=2},\phi_{\Delta S=2})$ plane, as defined in \eqref{eq:M12NP}, is displayed by the orange area in Fig.\ \ref{fig:eps-1}. We will now study how this constraint translates into the $Br(\kpn)-Br(\klpn)$, provided that the universality assumption \eqref{eq:uni} holds.

While in the scenario under consideration, the simple relation
$\phi_{\Delta S=2} = \phi_{K\to\pi\nu\bar\nu}$ holds by assumption, the same is  in general not true for the NP amplitudes $R_{\Delta S=2}$ and $R_{K\to\pi\nu\bar\nu}$. In order to measure the relative strength of $\Delta S=2$ and $\Delta S=1$ transitions, we therefore introduce the ratio
\be\label{eq:epsilon}
\epsilon = \frac{R_{K\to\pi\nu\bar\nu}}{R_{\Delta S=2}}\,.
\ee
 Thus in the scenario in question, the NP effects in $K^0-\bar K^0$ mixing and the $K\to\pi\nu\bar\nu$ decays are described by the three independent parameters $R_{\Delta S=2}$, $\phi$ and $\epsilon$. While $R_{\Delta S=2}$ and $\phi$ are severely constrained by the data on $\eps_K$, see Fig.\ \ref{fig:eps-1},  generally nothing can be said about the size of $\epsilon$. The larger $\epsilon$, the bigger is the NP effect on $K\to\pi\nu\bar\nu$ relative to its effect on $\Delta S=2$ observables.

While in specific scenarios in which NP enters at very different scales, its effects in 
$\Delta S=1$ transitions can significantly dominate over the ones in $\Delta S=2$ transitions \cite{Isidori:2006qy}, in the case of only one relevant NP scale it is natural to assume $\epsilon\sim\ord(1)$, i.\,e.\ that the influence of NP is roughly of equal size in $\Delta S=2$ and $\Delta S=1$ observables. 
Therefore in order to quantify the dependence of the actual size of $\epsilon$, we will consider different cases for its size.

\begin{figure}
\epsfig{file=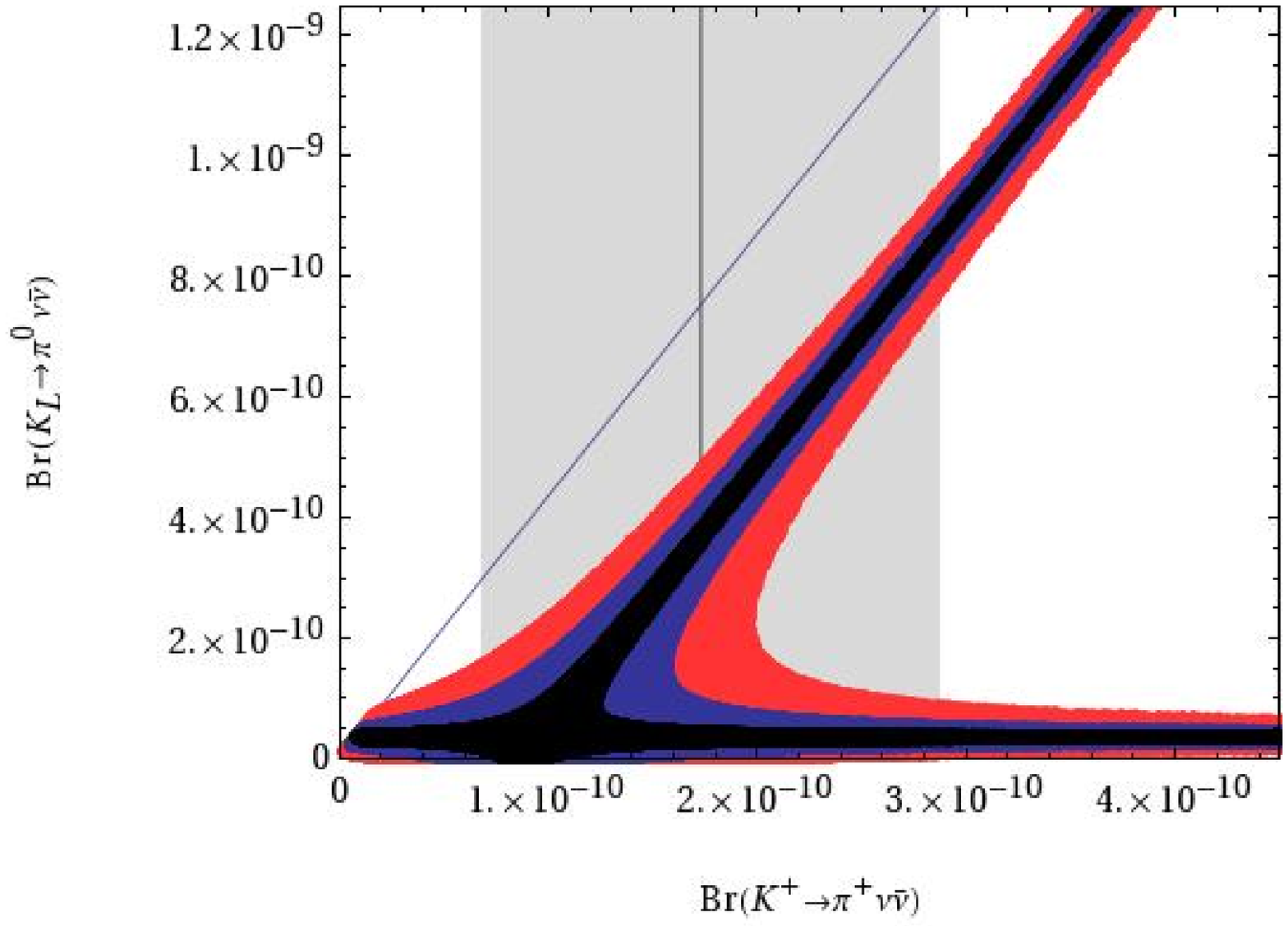,scale=.4}
\caption{ $Br(\klpn)$ as a function of $Br(\kpn)$ in the scenario of universal phases, for different values of $\epsilon$, and assuming a NP contribution to $\eps_K$ of at most $\pm 5\%$ of the SM. Black: $\epsilon=1$, blue (dark grey): $\epsilon=2$, red (lighter grey): $\epsilon=3$. The thin blue line shows the GN-bound, while the experimental $1\sigma$ range for $Br(\kpn)$ is displayed by the light grey band.\label{fig:e=123}}
\end{figure}

In Fig.\ \ref{fig:e=123} we show the implication of the $\eps_K$ constraint on the $Br(\kpn)-Br(\klpn)$ plane. To this end we scan over the parameters $R_{\Delta S=2}$ and $\phi$ and allow $\eps_K$ to deviate by at most $\pm 5\%$ from its SM prediction. The result of this scan is shown in Fig \ref{fig:eps-1}. Once the allowed ranges for $R_{\Delta S=2}$ and $\phi$ are fixed, the only degree of freedom entering the $K\to\pi\nu\bar\nu$ decays is the parameter $\epsilon$ in \eqref{eq:epsilon}, that measures the relative size of NP contributions to $K^0-\bar K^0$ mixing and to the $K\to\pi\nu\bar\nu$ decays. We study its impact on the observed correlation by considering three scenarios: $\epsilon=1$ (black region in Fig.\ \ref{fig:e=123}), $\epsilon=2$ (blue region) and $\epsilon=3$ (red region).
We clearly see again the two branches of Fig.\ \ref{fig:branches}, although the straight lines got broadened by the uncertainty in the $\eps_K$ constraint, and depending on the value of $\epsilon$ chosen. As could be expected the largest allowed range appears in the case $\epsilon=3$, as here the NP effects are largest and the $\eps_K$ constraint is least severe.
Generally, while in the vicinity of the SM value the strict correlation is diluted for $\epsilon > 1$, in case of large deviations from the SM the two branches appear still narrow and well separated from each other.

Having at hand these findings and the experimental result for $Br(\kpn)$ in \eqref{eq:kpnex}, we can deduce the {\em rough} upper bound
\be\label{eq:bound}
Br(\klpn)\lsim 9\cdot 10^{-10}
\ee
within the scenario in question. While this bound is by roughly $30\%$ stronger than the model-independent GN-bound, in contrast to the latter it depends on the additional assumptions discussed above. Therefore, in contrast to the GN-bound, the bound in eq.\ \eqref{eq:bound} is valid only in modelsthat predict equal $\Delta S=2$ and $\Delta S=1$ phases and NP contributions that are comparable in size.

We note that even more restrictive bounds can be obtained in specific models that predict in addition a non-trivial correlation between $K$ and $B$ physics, as e.\,g.\ in models with universally enhanced electroweak penguin contributions considered in \cite{Buras:2004ub} or in models with constrained minimal flavour violation \cite{Bobeth:2005ck,Haisch:2007ia}.

\subsection{\boldmath Sizable NP Contributions to $\eps_K$}\label{sec:largeNP}

While the experimentally observed amount of CP-violation in $K^0-\bar K^0$ mixing appears to be in rather good agreement with the SM prediction, recent studies \cite{Buras:2008nn,Buras:2009pj,Lunghi:2008aa} hint at the possibility that the SM alone cannot account for the full amount of CP-violation in this system, but that a NP contribution of roughly $+20\%$ is required to account for the data.
 While the situation is certainly not conclusive at present, improved determinations of $\hat B_K$ and the CKM parameters $|V_{cb}|$ and $\gamma$ will tell us whether indeed $(\eps_K)_\text{SM}$ is smaller than the data. Consequently, once the $K\to\pi\nu\bar\nu$ branching ratios will be determined experimentally, we will know whether the NP effects in $\eps_K$ are small, as analysed in Section \ref{sec:smallNP}, or sizable, as analysed in what follows. 

Therefore, instead of allowing $\eps_K$ to deviate by at most $\pm 5\%$ from its SM prediction, as done in Section \ref{sec:smallNP}, we will now assume a $+(20\pm5)\%$ NP contribution to $\eps_K$. The numerical analysis is then performed in an analogous way as in Section \ref{sec:smallNP}: the allowed ranges in the $(R_{\Delta S=2},\phi_{\Delta S=2})$ parameter space are determined from constraining
\be
\im (M_{12}^K)_\text{NP} = (20\pm5)\% \cdot \im (M_{12}^K)_\text{SM}\,.
\ee
The result is displayed by the light blue bands in Fig.\ \ref{fig:eps-1}.
In order to analyse the impact of this constraint on the $Br(\kpn)-Br(\klpn)$ plane, we again consider various scenarios for the parameter $\epsilon$, defined in \eqref{eq:epsilon}: 
 $\epsilon=0.5$ (purple region), $\epsilon=1$ (black region) and $\epsilon=2$ (blue region).

\begin{figure}
\epsfig{file=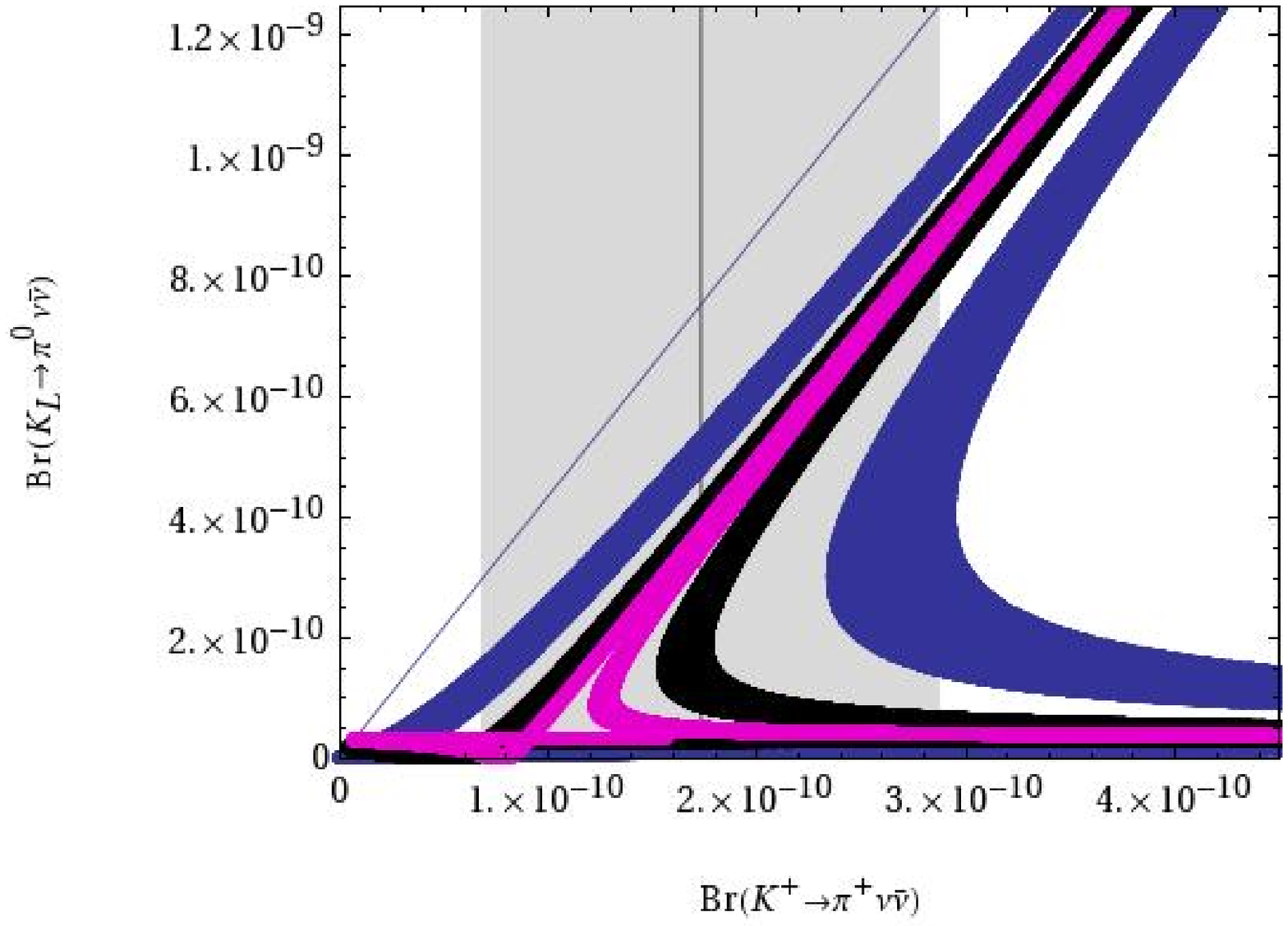,scale=.4}
\caption{$Br(\klpn)$ as a function of $Br(\kpn)$, for different values of $\epsilon$, and assuming a $+(20\pm5)\%$ NP contribution in $\eps_K$. Purple (lighter grey): $\epsilon=0.5$, black: $\epsilon=1$, blue (dark grey): $\epsilon=2$. The thin blue line shows the GN-bound, while the experimental $1\sigma$ range is displayed by the light grey band.\label{fig:BGcase}}
\end{figure}

In Fig.\ \ref{fig:BGcase} we show the constraints obtained on the $K\to\pi\nu\bar\nu$ decays in the scenarios in question.
We find that the two branches observed previously now split up into four sub-branches, moving further away from the branches in Fig.\ \ref{fig:branches} with increasing values of $\epsilon$. While for $\epsilon \le 1$ the stringent correlation between $\kpn$ and $\klpn$ is still maintained, for $\epsilon>1$ the allowed range in the $Br(\kpn)-Br(\klpn)$ plane  quickly moves away from these branches, so that the stringent model-independent correlation gets lost in that case.

While it seems at first sight that these findings weaken the power of the correlation analysed, one should keep in mind that sufficiently precise measurements of $\kpn$ and in particular $\klpn$ will  not be available within the next few years. The situation in $\eps_K$ on the other hand is much more promising, so that we will already know whether $\eps_K \simeq (\eps_K)_\text{SM}$ is satisfied in nature with high accuracy when the experimental situation in the $K\to\pi\nu\bar\nu$ system can finally help us to disentangle the NP flavour structure. So from today's point of view, basically two future scenarios are possible:
\begin{enumerate}
\item
We will know at that stage that $\eps_K \simeq (\eps_K)_\text{SM}$ with high precision. Then the measurement of both $K\to\pi\nu\bar\nu$ decay rates will show whether the correlation is satisfied in nature, and will thus provide a powerful test of the assumption of universal weak phases in $\Delta S=2$ and $\Delta S=1$ transitions.
\item
It will turn out that $\eps_K > (\eps_K)_\text{SM}$ so that the difference has to be accounted for by NP. Then the correlation in question can not be used to completely rule out the assumption of universal weak phases, but can on the other hand give precise information on the relative size of NP effects in $\Delta S=2$ and $\Delta S=1$ transitions within this specific scenario.
\end{enumerate}

\boldmath
\section{Non-Universal Phases in $\Delta S=2$ and $\Delta S=1$}
\unboldmath\label{sec:nonuniphase}

Let us now go beyond the simple assumption of universal weak phases in $\Delta S=2$ and $\Delta S=1$ processes and allow $\phi_{\Delta S=2}$ and $\phi_{K\to\pi\nu\bar\nu}$ to differ from each other. Clearly, if we abandon any correlation between $\Delta S=2$ and $\Delta S=1$ NP contributions and treat in particular their phases $\phi_{\Delta S=2}$ and $\phi_{K\to\pi\nu\bar\nu}$ as completely independent of each other, the constraint from $\eps_K$ has no more power to restrict the possible ranges in the 
$Br(\kpn)-Br(\klpn)$ plane. In this general case all values of $Br(\kpn)$ and $Br(\klpn)$ consistent with the GN-bound are possible.

However one should bear in mind that in most NP scenarios  $K^0-\bar K^0$ mixing and rare $K$ decays cannot be considered as completely independent of each other, as the former is induced by a double $(s\to d)\otimes (s\to d)$ transition, while the latter requires a single $(s\to d)$ transition.
While in the case of universal weak phases, $\Delta S=2$ and $\Delta S=1$ transitions were induced by only one new source of flavour violation, generally more than one contribution to each of these processes is present. If these contributions come along with independent weak phases, the resulting phases governing CP-violation in $K^0-\bar K^0$ mixing and in the rare $K$ decays are generally different from each other. However, in the case that the NP contributions display a similar hierarchy in both $\Delta S=2$ and $\Delta S=1$ systems, i.\,e. one dominating over the others, the universality relation  \eqref{eq:uni} is only weakly violated.
Therefore the aim of the present section is to quantify  how stable the correlation in the $K\to\pi\nu\bar\nu$ system discussed in Section \ref{sec:smallNP} is against small deviations from the universality assumption \eqref{eq:uni}.

In order to achieve this, we model the present case by setting
\be
R_{K\to\pi\nu\bar\nu} e^{i\phi_{K\to\pi\nu\bar\nu}}
= R_{\Delta S=2}  e^{i\phi_{\Delta S=2} } + P e^{i\psi}\,,
\ee
i.\,e.\ by splitting the NP contribution to $K\to\pi\nu\bar\nu$ into a part equal to the $\Delta S=2$ contribution and a second part parameterised by $P$ and $\psi$. 
The assumption of similar hierarchies in $\Delta S=2$ and $\Delta S=1$ is then fulfilled by the requirement $P\ll R_{\Delta S=2}$, while $0\le \psi<2\pi$ is chosen independently of $\phi_{\Delta S=2}$. Clearly in the limit $P\to 0$ the case of a universal phase with $\epsilon=1$ is recovered. 

As in Section \ref{sec:smallNP} we consider again the case that $\eps_K$ receives an at most $\pm5\%$ correction from NP. Thus the constraint on the $(R_{\Delta S=2},\phi_{\Delta S=2})$ parameter space is the same as in Section \ref{sec:smallNP} and displayed by the orange region in Fig.\ \ref{fig:eps-1}. When analysing the impact of this constraint on the $K\to\pi\nu\bar\nu$ system, we have now two free parameters, namely $P$ and $\psi$.

\begin{figure}
\epsfig{file=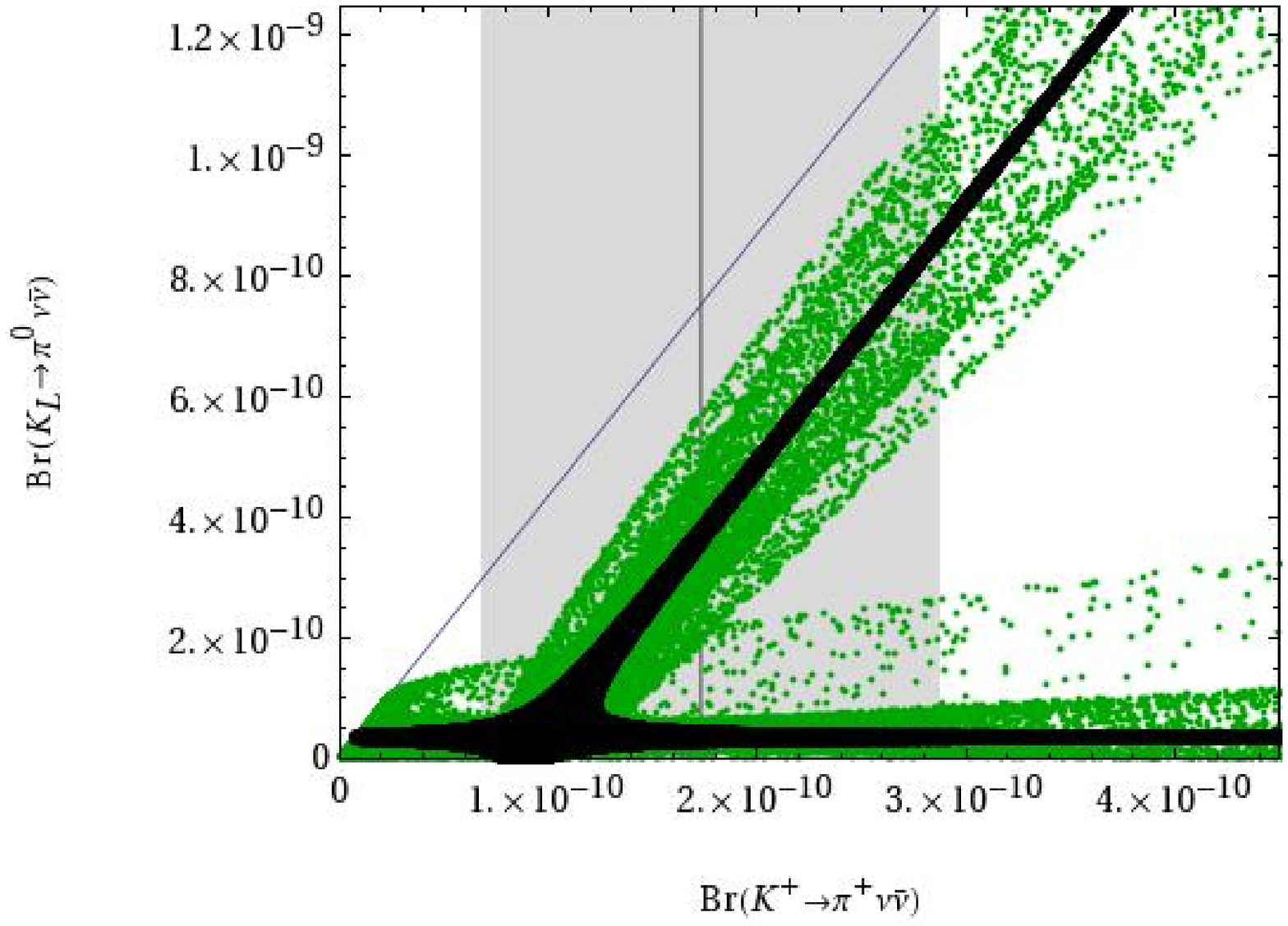,scale=.4}
\caption{$Br(\klpn)$ as a function of $Br(\kpn)$, assuming a NP contribution to $\eps_K$ of at most $\pm 5\%$ of the SM. 
Black: $P=0$, i.\,e. the NP phase in $\Delta F=2$ and $\Delta F=1$ are assumed to be equal. Green points: $P = 0.2 R_{\Delta S=2}$ and $0\le\psi<2\pi$ chosen randomly, breaking the universality of NP phases in $\Delta F=2$ and $\Delta F=1$.
The blue line shows the GN-bound, while the experimental $1\sigma$ range is displayed by the grey band.\label{fig:P}}
\end{figure}

In Fig.\ \ref{fig:P} we show the case $P = 0.2 R_{\Delta S=2}$ and the phase $\psi$ varied randomly between $0$ and $2\pi$.  This corresponds to $\phi_{\Delta S=2}$ and  $\phi_{K\to\pi\nu\bar\nu}$ differing by at most $\sim10^\circ$. 
We observe that even in this well-restricted scenario the two branches observed in the case of universal phases now broaden significantly, although they can still be distinguished from each other. Interestingly, this effect is largest in the case of large deviations from the SM, i.\,e. when $\klpn$ and/or $\kpn$ are strongly enhanced over their SM prediction. Clearly, if we allow for larger $P$ the correlation becomes even weaker and is completely lost already for $P\simeq0.35 R_{\Delta S=2}$, corresponding to a $\sim 20^\circ$ difference in $\phi_{\Delta S=2}$ and  $\phi_{K\to\pi\nu\bar\nu}$.

These findings tell us two things:
\begin{enumerate}
\item
The correlation in question  depends crucially on the universality of weak phases in $\eps_K$ and in the  $K\to\pi\nu\bar\nu$ decays. Already if the phases differ by $10^\circ$, the correlation is significantly weakend, albeit still visible, and completely lost once the weak phases $\phi_{\Delta S=2}$ and  $\phi_{K\to\pi\nu\bar\nu}$ differ by more than $20^\circ$.
\item
On the other hand, if in a model with a priori arbitrary weak phases a stringent correlation in the $Br(\kpn)-Br(\klpn)$ plane is observed, we know immediately that the weak phases entering $K^0-\bar K^0$ mixing and rare $K$ decays must be very close to equal. This happens indeed in the LHT model \cite{Blanke:2006eb,LHTupdate}, one of the most prominent representatives of the class of models with new sources of flavour and CP-violation but only SM operators.
\end{enumerate}

Let us mention the case of a non-SM-like $\eps_K$. As in the former case of universal weak phases and $\epsilon \le 1$, we have checked that the impact of a relevant NP contribution to $\eps_K$ is minor, in particular as the strict correlation between the NP phases in $\Delta S=2$ and $\Delta S=1$ transitions is partly washed out anyway in the present NP scenario.

Last but not least we note that the requirement of similar hierarchies in $\Delta S=2$ and $\Delta S=1$ amplitudes is more generally also fulfilled for 
\be
R_{K\to\pi\nu\bar\nu} e^{i\phi_{K\to\pi\nu\bar\nu}}
= \epsilon R_{\Delta S=2}  e^{i\phi_{\Delta S=2} } + P e^{i\psi}
\ee
for arbitrary $\epsilon$, provided $P \ll \epsilon R_{\Delta S=2}$ is satisfied. It is easy to see that such parameterisation effectively results in combining the results of this and the previous section. In particular for $\epsilon > 1$ the correlation in question would be further weakened. Therefore in a model like the LHT model, where a strong correlation in the $K\to\pi\nu\bar\nu$ system is observed, it is unlikely that NP affects more strongly the rare $K$ decays than the process of $K^0-\bar K^0$ mixing.

\section{The NP Operator Structure}\label{sec:newop}

\subsection{Preliminaries}

After the model-independent considerations of Sections \ref{sec:uniphase} and \ref{sec:nonuniphase}, we will now consider various possibilities for the NP operator structure and discuss the $K\to\pi\nu\bar\nu$ system and its possible correlation to $\eps_K$ in these scenarios. Specifically we discuss various NP prototypes: models in which FCNC processes are mediated by SM operators only, models in which NP dominantly induces right-handed currents and the only new $\Delta S=2$ operator is the $(V+A)\otimes(V+A)$ one, and models which induce left- and right-handed and/or scalar currents and thus the chirally enhanced left-right $\Delta S=2$ operators are present.

\subsection{Models without New Flavour Violating Operators}

In this section we consider models with arbitrary new sources of flavour and CP-violation, but only SM operators mediating FCNCs. A prime representative of this class of models is the Littlest Higgs model with T-parity (LHT), as discussed in detail in \cite{Blanke:2006sb,Blanke:2006eb}. 

As we have found in the previous section, already a relatively small deviation from the universality of CP-phases \eqref{eq:uni} of $\ord(10^\circ)$ significantly dilutes the correlation in the $K\to\pi\nu\bar\nu$ system implied by the $\eps_K$ constraint. Therefore without any additional knowledge on the flavour structure of a given NP model of this class, we cannot restrict the allowed ranges in the $K\to\pi\nu\bar\nu$ system, and one is tempted to think that in such kind of models no correlation between $Br(\kpn)$ and $Br(\klpn)$ is visible.

However the findings in the LHT model, that can be considered as a prototype of this class of models, tell us a different story. Also in that model, in spite of the presence of several independent weak phases, the stringent 2-branch-correlation, that we have encountered in Section \ref{sec:smallNP}, exists \cite{Blanke:2006eb,LHTupdate}. This in turn leads us to the conclusion that even in this more general scenario the weak phases in $\Delta S=2$ and $\Delta S=1$ transitions are in fact quite strongly correlated. From the discussion in the previous section, we can thus expect that in spite of various NP contributions entering $\Delta S=2$ and $\Delta S=1$ physics, the same contribution is dominant in both cases.

Having a closer look at the flavour structure of the LHT model and at the formulae describing $K^0-\bar K^0$ mixing and the $K\to\pi\nu\bar\nu$ decays within this model \cite{Blanke:2006eb,Goto:2008fj,LHTupdate}, we see that there are two independent contributions entering rare $K$ decays, of the structure
\be
\xi_2^{(K)} f_1\left((m_H^2)^2-(m_H^1)^2\right)\,,\quad \xi_3^{(K)} f_1\left((m_H^3)^2-(m_H^1)^2\right)\,.
\ee
Here $\xi_i^{(K)} = {V_{Hd}^{is}}^* V_{Hd}^{id}$, and $V_{Hd}$ is the new mixing matrix parameterising the mirror quark interactions with the usual SM quarks.
Furthermore $f_1$ is a loop function induced by the mirror quarks and heavy gauge bosons being exchanged in $Z$-penguin and box diagrams. As $f_1$ grows with increasing mirror quark mass splitting $(m_H^i)^2-(m_H^1)^2$, for $m_H^1 <m_H^2 < m_H^3$  the second contribution dominates over the first one, unless the $\xi_i^{(K)}$ exhibit a special hierarchy.

On the other hand, there are three distinct LHT contributions to $\Delta S=2$ physics, that can be described by
\begin{gather}
{\xi_2^{(K)}}^2 f_2\left((m_H^2)^2-(m_H^1)^2\right)\,,\\
{\xi_3^{(K)}}^2 f_2\left((m_H^3)^2-(m_H^1)^2\right)\,,\\
 {\xi_2^{(K)}}{\xi_3^{(K)}} \tilde f_2\left((m_H^2)^2-(m_H^1)^2,(m_H^3)^2-(m_H^1)^2\right)\,,
\end{gather}
where $f_2, \tilde f_2$ are the loop functions emerging from mirror quarks and heavy gauge bosons running in the  $\Delta S=2$ box diagrams. Also here in the case of non-degenerate mirror quark masses\footnote{Note that in the case of two quasi-degenerate mirror quark generations, $m_H^1\simeq m_H^2$, effectively only one contribution is present in both $\Delta S=2$ and $\Delta S=1$ transitions.}, $m_H^1 <m_H^2 < m_H^3$, 
one obtains a clear hierarchy between the various $\Delta S=2$ contribution, and the term proportional to ${\xi_3^{(K)}}^2$ turns out to be generally dominant.

Altogether we thus find that in spite of a priori various contributions to $\Delta S=2$ and $\Delta S=1$ processes in the LHT model, the same contribution, characterised by the largest mass splitting in the mirror quark sector, is dominant. As the loop functions $f_1$ and $f_2$ are real and flavour universal, CP-violating phases enter only through the ${\xi_3^{(K)}}$, ${\xi_3^{(K)}}^2$ factors in front. Therefore the relation $\phi_{\Delta S=2} = \phi_{K\to\pi\nu\bar\nu}$ is indeed satisfied with good approximation in the LHT model, and the strict correlation in the $K\to\pi\nu\bar\nu$ system can be understood.

\subsection{NP Inducing only Right-Handed Currents}

After discussing an explicit example of a model with new sources of flavour and CP-violation but no new operators, we now turn our attention to a simple scenario in which NP dominantly induces right-handed $(V+A)$ operators in addition to the SM left-handed ones. Such a scenario could emerge for instance in models with a heavy $Z'$ whose flavour violating couplings are purely right-handed, or in models in which the SM $Z$ boson couplings to right-handed quarks become flavour violating. Note however that at this stage we assume that the NP right-handed currents can {\em not} generate the chirally enhanced left-right operators contributing to $K^0-\bar K^0$ mixing. 
Specifically we consider the following structure for the effective $\Delta S=2$ and $\Delta S=1$ Hamiltonians:
\bea
\Heff(\Delta S=2) &=& C_\text{SM}^{\Delta S=2} (\bar sd)_{V-A}(\bar sd)_{V-A} \nn\\
&&{} + C_\text{NP}^{\Delta S=2} (\bar sd)_{V+A}(\bar sd)_{V+A}\,,\\
\Heff(\Delta S=1) &=& \Big[ C_\text{SM}^{\Delta S=1} (\bar sd)_{V-A} \nn\\
&&{} + C_\text{NP}^{\Delta S=1} (\bar sd)_{V+A}\Big] (\bar\nu\nu)_{V-A}\,.
\eea

As QCD is a non-chiral theory, the matrix elements of the operators $(\bar sd)_{V-A}(\bar sd)_{V-A}$ and $(\bar sd)_{V+A}(\bar sd)_{V+A}$ are equal, and we find  that the SM and NP contributions to $M_{12}^K$ are simply additive,
\be
M_{12}^K \propto C_\text{SM}^{\Delta S=2} + C_\text{NP}^{\Delta S=2}\,.
\ee
Similarly, as the $K\to\pi\nu\bar\nu$ decays are sensitive only to the vectorial $(\bar sd)_V$ current, also in this case the SM and NP parts are additive.

Altogether thus, the situation is completely analogous to the case of NP scenarios with only SM operators. Consequently, also in the present case, the correlation in the $K\to\pi\nu\bar\nu$ system can be used to test the universality of the phases of $ C_\text{NP}^{\Delta S=2}$ and $ C_\text{NP}^{\Delta S=1}$, and the previously made statements apply to this case. We note though that in specific NP models the allowed room in the $K\to\pi\nu\bar\nu$ plane can be further restricted by other $\Delta F=1$ constraints, in particular from $B$ decays. In order to keep our analysis as model-independent as possible, we do however not consider such additional constraints here.

On the other hand, our findings show that combining the data on $\eps_K$ with the data on the $K\to\pi\nu\bar\nu$ branching ratios can not help to distinguish this class of models from the case with no SM operators. Additional information from other decays is required. In fact, in \cite{Blanke:2008yr} it has been found that the correlation between $Br(\kpn)$ and the short distance contribution to $Br(K_L\to\mu^+\mu^-)$ offers an excellent probe of the handedness of new flavour violating currents. While the $K\to\pi\nu\bar\nu$ decays are sensitive to the vector part of the current, $(\bar sd)_V$, the $K_L\to\mu^+\mu^-$ mode measures its axial component, $(\bar sd)_A$. Therefore while in models with only SM operators a linear correlation between the two branching ratios is found \cite{LHTupdate}, in models with right-handed NP contributions the correlation between $Br(\kpn)$ and $Br(K_L\to\mu^+\mu^-)$ is an inverse one, as observed in the context of the custodially protected RS model \cite{Blanke:2008yr}. We note that although in the latter model  the tree level flavour changing $Z$ coupling to right-handed down-type quarks dominates the rare $K$ decays in question, this model does {\em not} belong to the class of models discussed in this section, as $K^0-\bar K^0$ mixing is dominated by tree level exchanges of KK gluons that sizably affect the chirally enhanced left-right operators \cite{Csaki:2008zd,Blanke:2008zb,Agashe:2008uz}. A detailed description of the custodially protected RS model, including a set of Feynman rules relevant for the study of FCNC processes, can be found in \cite{Albrecht:2009xr}.

\subsection{NP Inducing Left- and Right-Handed or Scalar Currents}

Finally let us briefly consider how our results change in the presence of NP left- and right-handed or scalar currents contributing to FCNC processes.
This happens for instance in a general MSSM  or in models with bulk fields in a warped extra dimension. It is common to both models that flavour violating effects can now also be mediated by right-handed currents, implying the presence of new operators beyond the left-handed SM ones.
In particular $K^0-\bar K^0$ mixing is then generally dominated by the left-right operators that receive strong chiral and QCD enhancements.
 The results of phenomenological analyses performed in the general MSSM \cite{Nir:1997tf,Isidori:2006qy} and in the RS model with custodial protection of flavour diagonal and non-diagonal $Zd_L^i\bar d_L^j$ couplings \cite{Blanke:2008yr}
let us anticipate that the correlation between $\klpn$ and $\kpn$ gets completely lost in that case, so that in principle every point in the $Br(\kpn)-Br(\klpn)$ plane consistent with the GN-bound can be reached.

It is not difficult to understand how this loss of correlation occurs. For the moment let us focus on the case of the custodially protected RS model; the situation in the general MSSM is similar albeit more complicated due to many different contributions competing with each other.

The dominant NP contribution to $\eps_K$ in the custodially protected RS model arises due to tree level exchanges of KK gluons that sizably affect the chirally enhanced left-right operators \cite{Csaki:2008zd,Blanke:2008zb,Agashe:2008uz}. As the Wilson coefficients of these operators are in general most severely costrained by the data \cite{Bona:2007vi}, the NP contribution to $\eps_K$ is thus fully dominated by the product of a left-handed and a right-handed transition.
The $K\to\pi\nu\bar\nu$ decays on the other hand, being $\Delta S=1$ transitions, can be mediated either only by a left- or by a right-handed transition, but not by a product of both, as is the case for the $\Delta S=2$ left-right operators, and the relevant $\Delta S=1$ amplitude is given by the sum of these contributions. We see immediately that different parts of the NP flavour sector enter $\eps_K$ and the $K\to\pi\nu\bar\nu$ decays, so that a correlation between the relevant weak phases $\phi_{\Delta S=2}$ and  $\phi_{K\to\pi\nu\bar\nu}$ cannot be expected. The result of the numerical analysis \cite{Blanke:2008yr} confirms these findings: No correlation between $Br(\kpn)$ and $Br(\klpn)$ appears.

A similar situation is to be expected in all models where flavour changing neutral currents are mediated by both left- and right-handed currents and/or by scalar currents. Due to their strong chiral and QCD enhancement, the induced left-right operator contribution will very likely dominate the NP contribution to $K^0-\bar K^0$ mixing, so that no correlation of weak phases in $\Delta S=2$ and $\Delta S=1$ processes appears. Consequently, the $\eps_K$ constraint can not be used to restrict the $K\to\pi\nu\bar\nu$ decay rates.

\section{Conclusions}
\label{sec:sum}

In the present paper we have studied the impact of the constraint from $\eps_K$ on the allowed range in the $Br(\kpn)-Br(\klpn)$ plane in various NP scenarios. The main messages from this analysis are as follows:
\begin{enumerate}
\item\label{it:univ}
In NP scenarios in which  a single new weak phase affects universally both $K^0-\bar K^0$ mixing and the $K\to\pi\nu\bar\nu$ decays, the experimental constraint from $\eps_K$ implies a strict correlation between the $\kpn$ and $\klpn$ decay rates. This correlation consists basically of two branches, one parallel to the $Br(\kpn)$ axis and one parallel to the GN-bound, and crossing each other in the SM prediction. While the  broadness of the observed branches depends on the experimental and theoretical error on $\eps_K$ as well as the relative size of NP contributions in $K^0-\bar K^0$ mixing and the $K\to\pi\nu\bar\nu$ decays, the general prediction appears to be stable against $\ord(1)$ modifications of the latter ratio.
\item
If the assumption of a universal new phase is relaxed, the above correlation gets softened, so that in the case of completely uncorrelated $\Delta S=2$ and $\Delta S=1$ NP phases, no visible correlation in the $Br(\kpn)-Br(\klpn)$ plane exists.
\item
The correlation in question is also softened if $\eps_K$ is affected by relevant NP contributions. In the scenario of universal weak phases its power will then be shifted towards giving precise information on the relative size of the NP amplitudes in $K^0-\bar K^0$ mixing and the $K\to\pi\nu\bar\nu$ system within the specific NP scenario of a universal $\Delta S=2$ and $\Delta S=1$ phase.
\item
On the other hand in many NP scenarios with only SM operators, with the LHT model being a famous example, it appears that even in the presence of more than one weak phase, $\Delta S=2$ and $\Delta S=1$ NP are correlated, leading in particular to roughly equal phases in $K^0-\bar K^0$ and  $K\to\pi\nu\bar\nu$. Then the correlation of \ref{it:univ}. is partially recovered, albeit not as stringent as in the case of strictly equal phases. 
\item 
The situation changes drastically once the chirally enhanced left-right operators are allowed to contribute to $K^0-\bar K^0$ mixing. As such a structure can not appear in the $K\to\pi\nu\bar\nu$ decays, it is natural to assume in this case completely independent phases  in these two processes, as the various flavour violating transitions enter $\Delta S=2$ and $\Delta S=1$ transitions in a very different manner. Consequently no correlation in the $K\to\pi\nu\bar\nu$ system appears.
\end{enumerate}

The $K\to\pi\nu\bar\nu$ decays serve as a unique probe of the NP flavour structure. The correlation observed and analysed in the present paper offers a powerful tool to test the universality of NP in $\Delta S=2$ and $\Delta S=1$ transitions. While it is possible to obtain independent phases already in NP scenarios with only SM operators, we argued that a very likely scenario for different phases entering  $\Delta S=2$ and $\Delta S=1$ transitions is the presence of new flavour violating operators contributing to $K^0-\bar K^0$ mixing, with the most plausible possibility being the chirally enhanced left-right operators contributing to $\eps_K$. In that sense the present analysis suggests to consider the observed correlation as a test of the operator structure of the NP flavour sector.

\subsection*{Acknowledgements}

I would like to thank Andrzej Buras and Paride Paradisi for very useful discussions and comments on the manuscript.
This research was partially supported by 
the DFG Cluster of Excellence `Origin and Structure of the Universe'.

\providecommand{\href}[2]{#2}\begingroup\raggedright\endgroup

\end{document}